\documentclass[%
 reprint,
 amsmath,amssymb,
 aps,
]{revtex4-1}

\usepackage{graphicx}
\usepackage{dcolumn}
\usepackage{bm}

\usepackage[T1]{fontenc}
\usepackage[latin9]{inputenc}
\usepackage{textcomp}
\usepackage{bashful}
\usepackage{soulpos}
\usepackage{soul}
\usepackage{relsize}
\usepackage{amsmath}

\newcommand{\TC}{{T_{\mathrm c}}}

\usepackage{epstopdf}
\AppendGraphicsExtensions{.tif}

\usepackage{fixltx2e}

\begin{document}

\preprint{APS/123-QED}

\title{Optical Receiver with Helicity Dependent Switching of Magnetization}

\author{Zubair Al Azim}
\email{zazim@purdue.edu}
\thanks{author contributed equally}
\affiliation{Department of Electrical and Computer Engineering, West Lafayette, IN 47907}

\author{Thomas A. Ostler}
\thanks{author contributed equally}
\affiliation{
 Faculty of Arts, Computing, Engineering and Sciences, Sheffield Hallam University, Howard Street, Sheffield, S1 1WB, UK
}
\affiliation{ Nanomat / Q-MAT / CESAM and European Theoretical Spectroscopy Facility, Universit\'e de Li\`ege, B-4000 Li\`ege, Belgium}
\author{Chudong Xu}
\affiliation{College of Electronic Engineering, South China Agricultural University, 
Guangzhou, Guangdong 510642, China
}

\author{Kaushik Roy}%
\affiliation{Department of Electrical, Computer, and Energy Engineering, West Lafayette, IN 47907}


\date{\today}

\begin{abstract}
In this work, we propose helicity-dependent switching (HDS) of magnetization of Co/Pt for energy efficient optical receiver. Designing a low power optical receiver for optical-to-electrical signal conversion has proven to be very challenging. Current day receivers use a photodiode that produces a photocurrent in response to input optical signals, and power hungry trans-impedance amplifiers are required to amplify the small photocurrents. Here, we propose light helicity induced switching of magnetization to overcome the requirement of photodiodes and subsequent trans-impedance amplification by sensing the change in magnetization with a magnetic tunnel junction (MTJ). Magnetization switching of a thin ferromagnet layer using circularly polarized laser pulses have recently been demonstrated which shows one-to-one correspondence between light helicity and the magnetization state. We propose to utilize this phenomena by using digital input dependent circularly polarized laser pulses to directly switch the magnetization state of a thin Co/Pt ferromagnet layer at the receiver. The Co/Pt layer is used as the free layer of an MTJ, the resistance of which is modified by the laser pulses. With the one-to-one dependence between input data and output magnetization state, the MTJ resistance is directly converted to digital output signal. Our device to circuit level simulation results indicate that, HDS based optical receiver consumes only $0.124$ $pJ/bit$ energy, which is much lower than existing techniques.
\end{abstract}

\maketitle


\section{\label{sec:level1}Introduction }

Optical interconnect is considered to be the leading candidate for off-chip communication in future multi-core systems due to its negligible channel loss and higher noise immunity~\cite{Biberman2012,Mishra2013}. However, in order to broaden its commercial application, optical interconnects must offer orders of magnitude higher energy efficiency compared to existing electrical interconnects~\cite{Krishnamoorthy2011}. Significant progress has been made in recent years to lower the energy consumption in optical interconnects, especially in the conversion of electrical to optical signals~\cite{Biberman2012}. Designing highly energy efficient receivers for optical-to-electrical signal conversion, however, remains a challenge. Present-day optical receivers need to convert small photocurrents to CMOS compatible voltage signals, which leads to several design challenges~\cite{Krishnamoorthy2011}. The direct use of optical signals to induce switching of magnetization can potentially overcome some of these challenges.

Magnetization reversal using only ultrafast laser pulses has recently been demonstrated in several experiments~\cite{Stanciu2007,Radu2011} and remains a topic of great interest~\cite{Li2015}. The demonstrations have shown magnetization switching can either be dependent or independent of the laser pulse helicity. Laser helicity dependent switching (HDS) is more desirable for the conversion of optical-to-electrical signal because of the inherent one-to-one correspondence between the optical signal and magnetization state. The reversal of magnetization through single-shot laser pulses has been shown to be helicity independent and a purely thermal process, which is observed mostly in ferrimagnets~\cite{Radu2011}. Although, single-shot switching was recently observed in ferromagnetic Pt/Co/Pt multilayer structures~\cite{Vomir2017}, the switching was shown to be helicity-independent and the time-scale of the process was on the order of nanoseconds. Exchange coupled ferromagnetic/ferrimagnetic ((Co/Pt)/GdFeCo) multilayers were also shown~\cite{Gorchon2017} to exhibit ultrafast switching (within $7$ $ps$), however, no helicity-dependence was shown. HDS was previously found to occur through the action of multiple laser pulses, though again, mostly in ferrimagnetic materials~\cite{Stanciu2007}. The necessity of using exotic ferrimagnetic materials is undesirable for the conversion of optical-to-electrical signal.

In this work, we propose helicity-dependent switching of magnetization in a thin Co/Pt ferromagnet layer for an energy efficient optical receiver. This switching process has recently been demonstrated experimentally in~\cite{Hadri2016,Lambert2014}. Laser pulses with right-hand circular polarization (RHCP or $\sigma+$) were shown to the reverse magnetization from a `down' to an `up' state and vice-versa. With the use of HDS, it becomes possible to have one-to-one correspondence between input data and output magnetization state. This can be achieved by transmitting laser pulses with opposite circular polarization (either right-hand or left-hand) for digital `0' or `1' input data. We should point out that multiple pulses are needed to switch the ferromagnet layer as shown in~\cite{Hadri2016,Lambert2014}. The Co/Pt layer can be used as the free layer of a magnetic tunnel junction (MTJ). Laser pulses modify the MTJ resistance in accordance to the helicity and this resistance change can be sensed through a resistive divider action. We will first present the modeling of HDS in ferromagnets using the Landau-Lifshitz-Bloch (LLB) formalism. The model is developed in-house and is outlined in~\cite{Ostler2015b,Ostler2014,Atkinson2016}. Next, we will discuss how we incorporate the magnetization dynamics with an MTJ resistance model in order to perform device to circuit level simulation. We will conclude by presenting the details of our proposed optical receiver and evaluating its performance.

\section{Modeling of HDS in Ferromagnets and Incorporating with Circuit Simulation}
\subsection{\label{sec:level2}Landau-Lifshitz-Bloch Model}
The LLB equation describes the time evolution of a magnetic macrospin. The equation allows for longitudinal relaxation (as well as transverse precessional and relaxation behaviour) of the magnetization, and was derived by Garanin~\cite{Garanin1997} within a Mean Field approximation from the classical Fokker-Planck equation for atomic spins interacting with a heat bath. In this sense the equation attempts to describe, in a spatially averaged way, the motion of an ensemble of magnetic moments. Models based on the resulting expressions have been shown to be consistent with atomistic spin dynamics simulation~\cite{Kazantseva2008}, as well as comparisons with experimental observations, for example, in laser induced demagnetization~\cite{Mendil2014}. The equation is similar to the Landau-Lifshitz-Gilbert (LLG) equation~\cite{Gilbert2004}, with precessional and relaxation terms, but with an extra term that deals with changes in the length of the magnetization:
\begin{eqnarray}
&&\mathbf{\dot{m}}_i=-\gamma \lbrack \mathbf{m}_i\times  \mathbf{H}^{%
\mathrm{eff}}_i ]+\frac{\gamma \alpha _{\parallel}}{m_i^{2}}\left(%
\mathbf{m}_i\cdot  \mathbf{H}^{\mathrm{eff}}_i\right) %
\mathbf{m}_i \nonumber \\
&&\qquad {}-\frac{\gamma \alpha
_{\perp}}{m_i^{2}} \left[\mathbf{m}_i\times \left \lbrack
\mathbf{m}_i\times \mathbf{H}_i^{\mathrm{eff}} %
\right] \right],
\label{eq:llb}
\end{eqnarray}
where $\mathbf{m}_i$ is the spin polarization, $\mathbf{M}_i/M_s(0)$. The spin polarization tends towards equilibrium, $m_e$, which is a temperature dependent quantity. ${\alpha_{\parallel}}$ and
$\alpha_{\perp}$ are dimensionless longitudinal and transverse damping
parameters. $\gamma$ is the gyromagnetic ratio taken to be the free electron value. The LLB equation is valid for finite temperatures and even above $\TC$, though the damping parameters and effective fields are different below and above $\TC$. For the transverse damping parameter:
\begin{equation}
\alpha_{\perp} = \begin{cases}
\lambda\Big( 1 - \frac{T}{3\TC} \Big) & {T< \TC} \\
\lambda \frac{2T}{3\TC} & {T\geq \TC}
\end{cases}
\label{eq:alphaperp}
\end{equation}
and for the longitudinal:
\begin{equation}
\alpha_{\parallel}  =  \lambda \frac{2T}{3\TC} \qquad {\text{for all $T$.}}
\label{eq:alphapar}
\end{equation}
For a single particle, the effective field $\mathbf{H}^{\mathrm{eff}}_i$ is given by \cite{Garanin1997}:  
\begin{equation}
  \mathbf{H}^{\mathrm{eff}}_i = \mathbf{B}+\mathbf{H}_{A,i}+
       \frac{1}{2\tilde{\chi}_{i,\Vert }}\left(1-\frac{m_i^{2}}{m_{i,\rm e}^{2}}\right) \mathbf{m}_i + \mathbf{H}_{\rm{e},i} + \mathbf{H}_{\rm{demag},i}
    \label{e:Heffm}
\end{equation}
where $\mathbf{B}$ represents an external magnetic field,
 $\mathbf{H}_{A,i}$  is the uniaxial easy axis anisotropy field and $\mathbf{H}_{\rm{e},i}$ is the exchange field. $\tilde{\chi}_{\parallel}$ is the parallel susceptibility which is defined by $\tilde{\chi}_{\parallel} = \partial m_{\parallel} / \partial H_{\parallel}$. The final term in equation~\ref{e:Heffm}, $\mathbf{H}_{\rm{demag},i}$ is the demagnetizing field. 

In the above equations, $\lambda$ is a microscopic parameter which
characterizes the coupling of the individual, atomistic spins with
the heat bath. We choose the value of $\lambda$ to be $0.025$ for this work, however, the demagnetization process is strongly dependent on this parameter. Table~\ref{table:tab1} shows a summary of the parameters that are used in our model.
\begin{table}%
\caption{\label{table:tab1}%
Physical parameters entering into the LLB model for Co/Pt}
\begin{ruledtabular}
\begin{tabular}{ccc}
\textrm{Quantity}&
\textrm{Value}&
\textrm{Units}\\
\colrule
$\lambda$ & $0.025$ & \\
$M_{s}(0)$ & $1.438 \times 10^{6}$ & $J T^{-1} m^{-3} $\\
$K(0)$ & $2.56 \times 10^{6}$ & $J m^{-3}$\\
$\gamma$ & $1.76 \times 10^{-11}$ & $T^{-1} s^{-1}$\\
System size & $100 \times 100$ & $nm^{2}$\\
No. of Macrospins & $50 \times 50 \times 1$ & \\
Macrospin Size & $2 \times 2 \times 0.6$ & $nm^{3}$\\
$T_{C}$ & $650$ & $K$\\
\end{tabular}
\end{ruledtabular}
\end{table}

To account for the laser heating in this model, we utilize the semi-classical two-temperature model~\cite{Chen2006,Atxitia2015a} of laser heating. This model defines a temperature associated with the electron and phonon heat baths through the simplified equations:
\begin{equation}
C_e \frac{\partial T_e(x,y)}{\partial t} = -G (T_e(x,y) - T_l(x,y)) + P(x,y,t) 
\end{equation}
\begin{equation}
C_l \frac{\partial T_l(x,y)}{\partial t} = G (T_e(x,y) - T_l(x,y)) + (T_l(x,y) - T_{eq})/{\tau_{c}}
\end{equation}
where $C_{e,l}$ and $T_{e,l}(x,y)$ are the electron (lattice) specific heats and temperatures, respectively, and $G$ is the electron-lattice coupling constant. $T_{eq}$ is the equilibrium temperature set to $300$ $K$ and $\tau_{\text{c}}$ is the cooling time, which we assume to be $100$ $ps$. The time-and-spatially dependent laser power is assumed to be Gaussian in both time and space:
\begin{eqnarray}
P(x,y,t)& =& \mathcal{F}\exp \Bigg(-\Bigg(\frac{t-t_0}{\tau_p}\Bigg)^2\Bigg)\\ \nonumber
&&\times\exp\Bigg(-\frac{(x-x_0)^2}{2\sigma_x^2}\Bigg)\exp\Bigg(-\frac{(y-y_0)^2}{2\sigma_y^2}\Bigg)
\label{eq:2DP}
\end{eqnarray}
where $t_0$ is the pump delay, $\tau_p$ is the pump width which we choose to be $50$ $fs$, $x_0$ and $y_0$ are the pump centers in $x$ and $y$ respectively, which are both set to $50$ $nm$, and $\sigma_{x,y}$ are the spatial widths in $x$ and $y$ which is set to $50$ $\mu m$ which is a typical width of a femtosecond laser experiment which essentially provides uniform heating to our element.

As well as implementing the spatial dependence of the pump fluence, we have also added a spatial dependence of the field intensity arising from the inverse Faraday effect in a phenomenological way (IFE, which signifies the generation of a magnetic field according to light polarization~\cite{Pershan1966}). The width of the IFE field temporally was chosen to be $9.5$ $ps$. The field amplitude from IFE was chosen to be $5$ $T$, the sign of which was altered in accordance with laser helicity. Considering the relatively short duration of the laser pulse, a temporal width of $9.5$ $ps$ is rather long given that the optical coherence time in metals should be comparable to the pulse duration. However, similar demagnetization times and degree of demagnetization/switching was observed experimentally in~\cite{Hadri2016,Lambert2014}. Furthermore, the amplitude of the field is somewhat difficult to quantify. In the theory of the IFE, the effect of the light is to induce a magnetization. Here, we assume that a phenomenological field gives rise to this change in magnetization, though this approximation has been used to good effect in previous works~\cite{Vahaplar2009} and remains an interesting and open question~\cite{John2017}. In~\cite{Hadri2016,Lambert2014}, helicity dependent switching in ferromagnet occurred through the action of multiple laser pulses to allow sufficient time for transfer of angular momentum from the laser to the magnet. In our model, we allow $250$ $ps$ time interval between successive laser pulses such that heating due to laser pulses do not randomize the magnetization. The values of IFE field width and duration as well as the successive pulse separation interval were chosen to roughly approximate the number of laser pulses required to induce switching in~\cite{Hadri2016,Lambert2014}. The size of our elements are much smaller than those in the experiments of~\cite{Hadri2016,Lambert2014}. Hence, our switching is completed (to saturation) faster than in experiments, as the effects from the demagnetizing field is much smaller. As our focus here is not to understand the origins of all-optical switching but pose a potential application of the phenomena, a complete one-to-one agreement of the theory and experiment is not necessary.
\begin{center}
\begin{figure}
\includegraphics[width=\columnwidth]{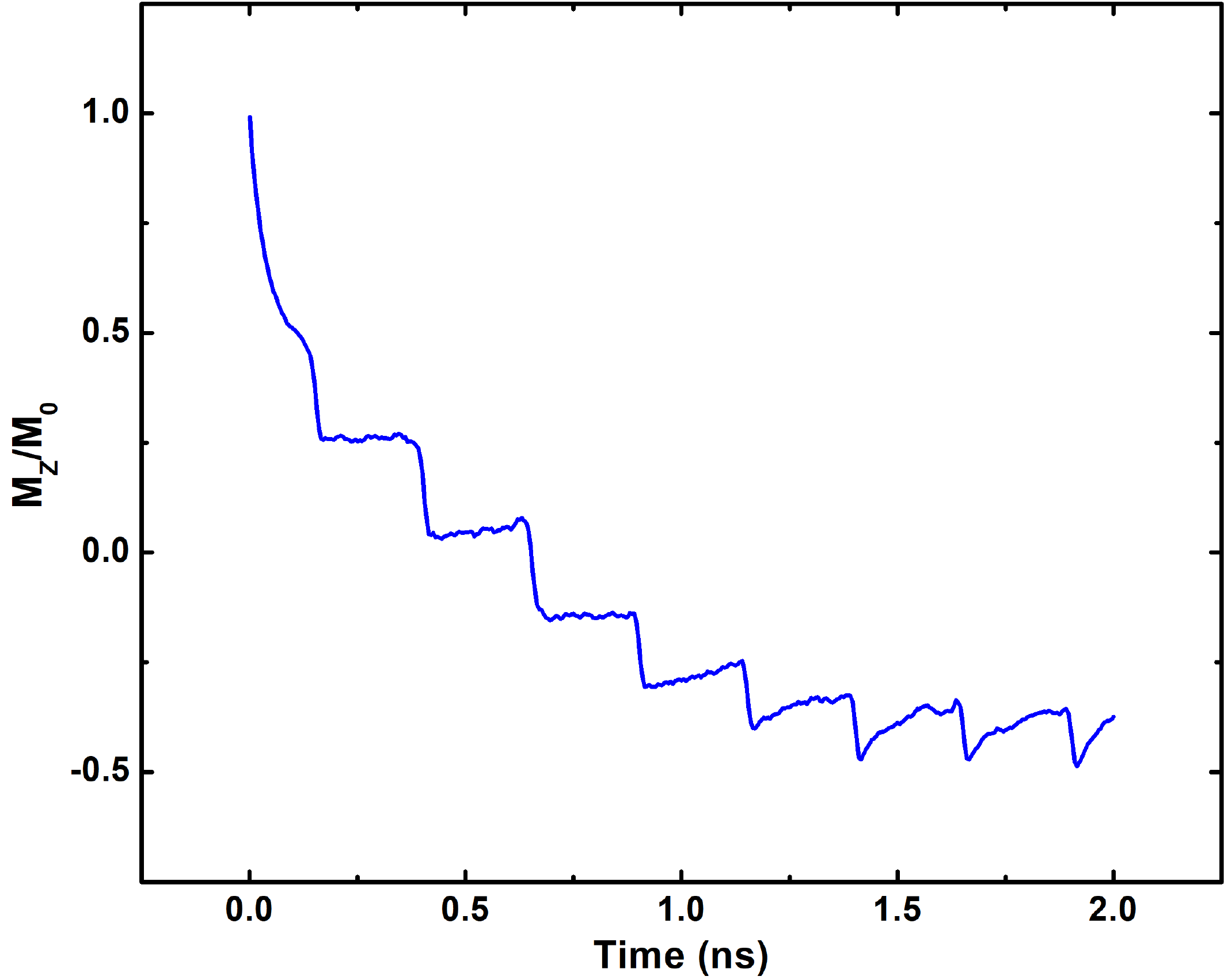}
\caption{Temporal response of a Co/Pt layer magnetization in response to LHCP($\sigma-$) laser pulses}
\label{fig:single}
\end{figure}
\end{center}

\begin{figure}
\centering
\includegraphics[width=\columnwidth]{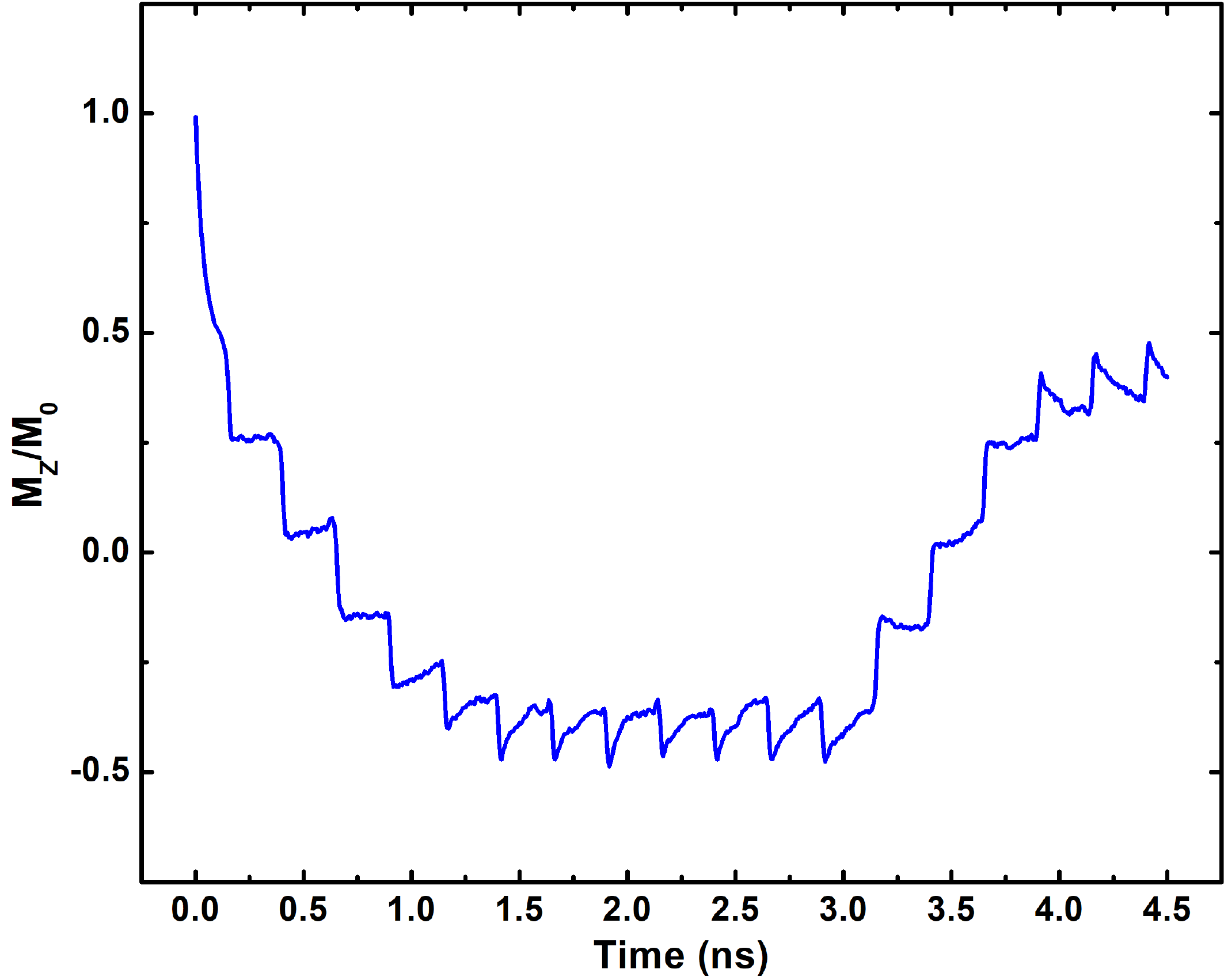}
\caption{Temporal response of a Co/Pt layer magnetization in response to both LHCP($\sigma-$) and RHCP($\sigma+$) laser pulses}
\label{fig:multi}
\end{figure}

\vspace{-2em}
In Fig.~\ref{fig:single}, we show the temporal variation of Co/Pt layer magnetization in response to left-hand circularly polarized (LHCP or $\sigma -$) laser pulses. The initial magnetization was taken to be pointing in the `up' direction ($M_{z}/M_{0} = +1$). The number of pulses required to reverse the magnetization is $6$ and the reversal takes $\sim1.4$ $ns$ as shown in Fig.~\ref{fig:single}. Note, that the degree of reversal is limited ($M_{z}/M_{0}$ saturates to $\sim-0.5$ in Fig.~\ref{fig:single}) because of the fact that the equilibrium (operating) temperature is kept fixed at room temperature. The temporal magnetization response to multiple helicity laser pulses is shown in Fig.~\ref{fig:multi}. Starting again from an initially `up' magnetized state, the magnetization reverses in $\sim1.4$ $ns$ in response to $\sigma -$ pulses. We continue to apply $\sigma -$ pulses upto $3$ $ns$. However, once the magnetization saturates, further application of $\sigma -$ pulses do not change the magnetization. After $3$ $ns$, the laser helicity is reversed to $\sigma +$. The application of $\sigma +$ pulses again reverses the magnetization towards `up' state as shown in Fig.~\ref{fig:multi}. This demonstrates the possibility of repeated operation by altering the laser helicity, which is necessary for the interconnect application.

\subsection{\label{sec:level3}Incorporating HDS with an MTJ for Circuit Analysis}
\begin{figure}
\centering
\includegraphics[width=\columnwidth]{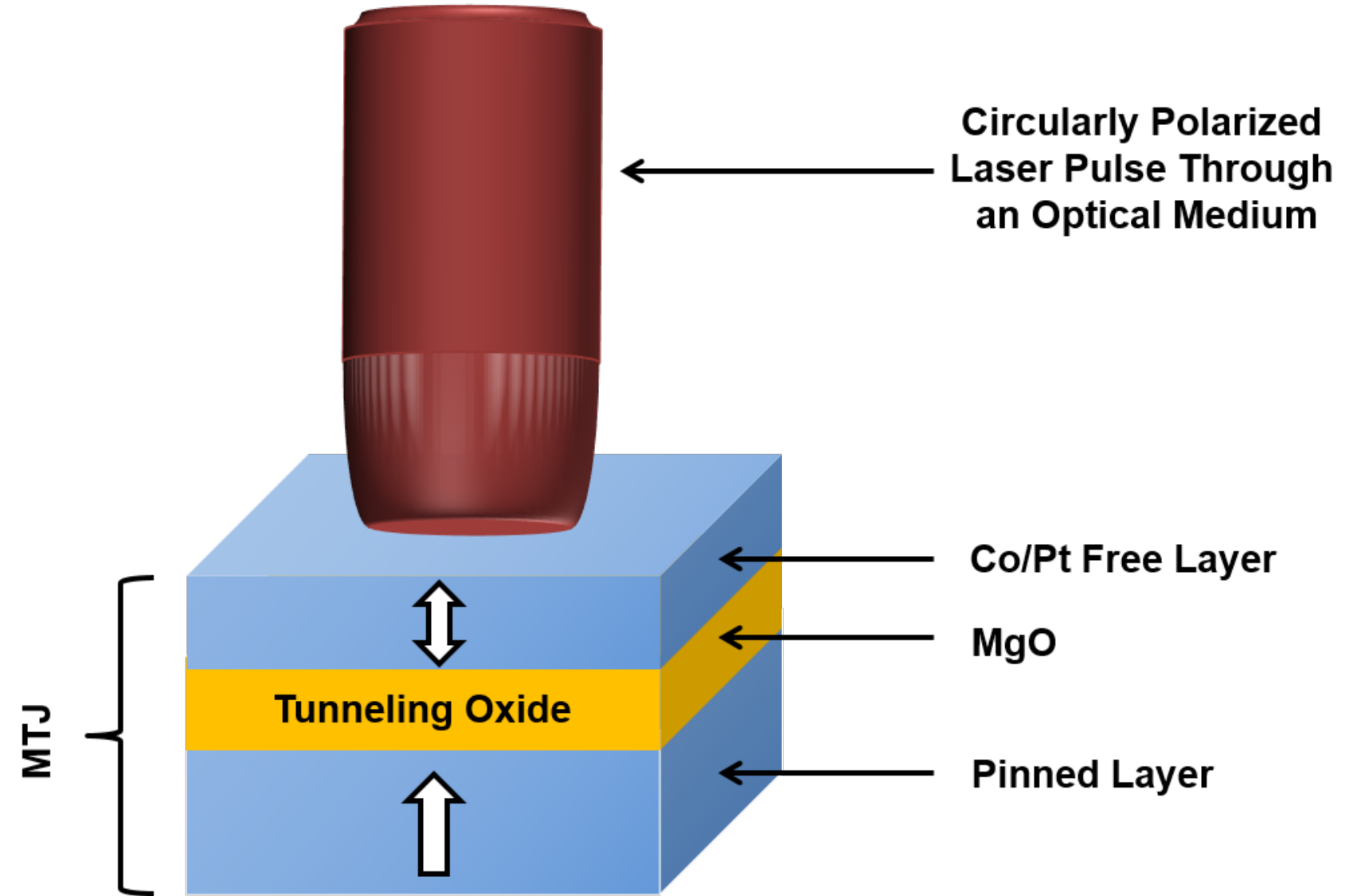}
\caption{Co/Pt as the free layer of an MTJ, the resistance of which can be tuned by using circularly polarized laser pulses}
\label{fig:device}
\end{figure}

In order to use HDS for circuit application, a thin Co/Pt layer is used as the free layer of an MTJ as shown in Fig.~\ref{fig:device}. The resistance of this MTJ is tuned by the laser helicity-induced magnetization control of the Co/Pt layer. With the direction of the MTJ pinned layer shown in Fig.~\ref{fig:device}, the MTJ resistance is high (R\textsubscript{AP}) when the Co/Pt magnetization is close to the `down' state and MTJ resistance is low (R\textsubscript{P}) when the Co/Pt magnetization is close to the `up' state. The resistance of the MTJ stack is modeled by non-equilibrium Green's Function (NEGF) formalism and abstracted into a behavioral MTJ resistance model. A detailed description of this method can be found in~\cite{Fong2011}. The laser induced magnetization data is incorporated with this behavioral MTJ resistance model to evaluate the laser helicity induced MTJ resistance change. The resistance of the MTJ is then subsequently integrated with $45$ $nm$ CMOS technology to evaluate the circuit operations.  

\section{Circuit Operation and Discussion}
\subsection{\label{sec:level4}Optical Receiver using HDS}
\begin{center}
\begin{figure}[!hbtp]
\includegraphics[width=\columnwidth]{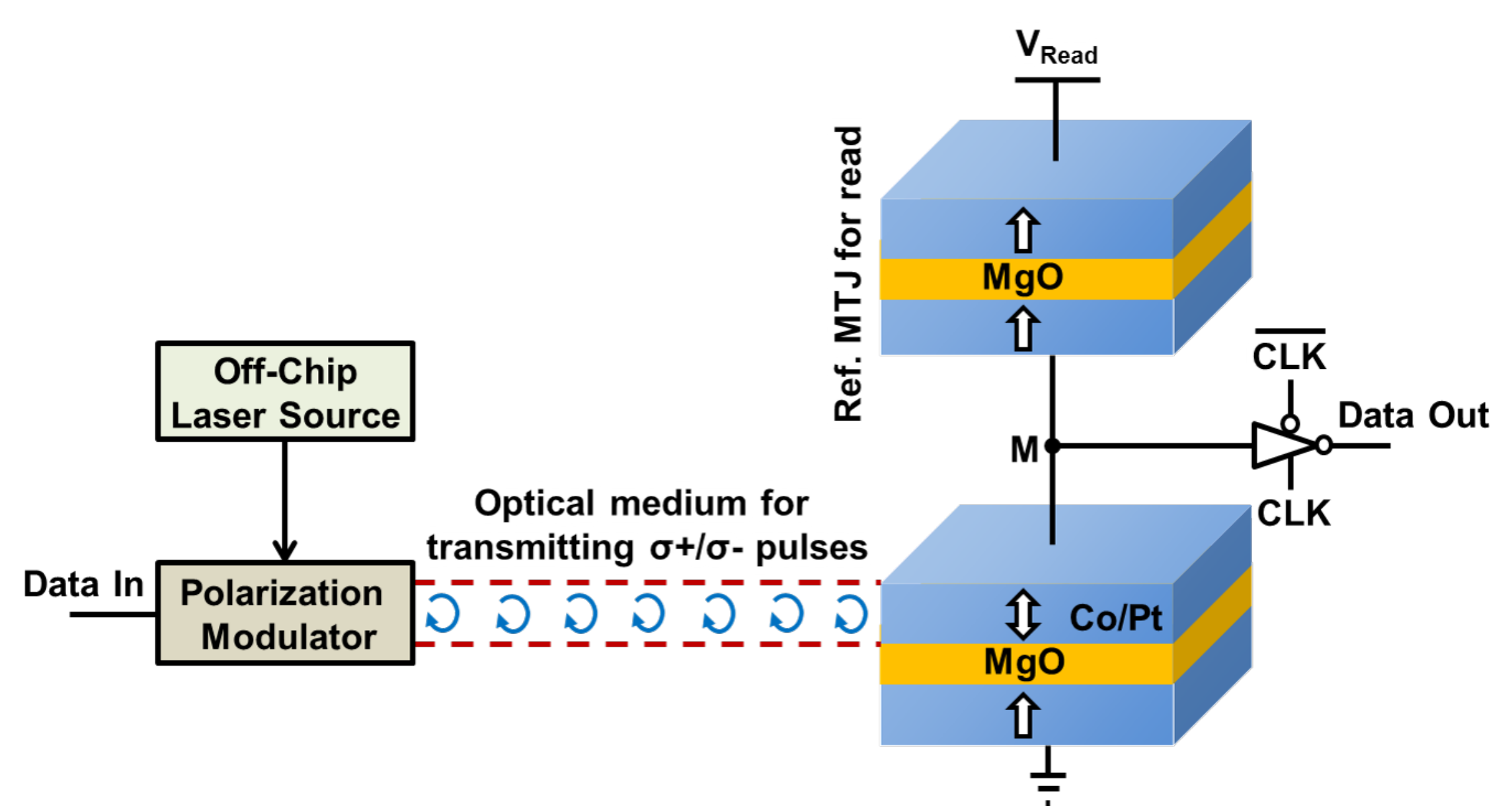}
\caption{Schematic of the optical interconnect scheme with HDS based receiver}
\label{fig:circuit_schematic}
\end{figure}
\end{center}

The schematic of the optical interconnect circuit using HDS at the receiver is shown in Fig.~\ref{fig:circuit_schematic}. As mentioned previously, the Co/Pt ferromagnet layer is used as the free layer of an MTJ at the receiver. The magnetization state of this Co/Pt layer is modified by using circularly polarized laser pulses. The change of the MTJ resistance is sensed by using the reference MTJ as shown in Fig.~\ref{fig:circuit_schematic}, which creates a resistance divider network. A read current is passed through the two MTJ resistances (connected in series) by using the terminal $V_{Read}$. The read current sets the voltage at node `M' in Fig.~\ref{fig:circuit_schematic} in accordance to the resistance of the bottom MTJ. This resistive divider MTJ network drives a clocked CMOS inverter as shown in Fig.~\ref{fig:circuit_schematic} to produce the appropriate digital output signal. A digital input data controls the laser polarization through the use of a binary circular polarization modulator~\cite{Abidin2012} at the input side. The optical modulator controls the helicity of the laser input from an off-chip laser source and transmits the resultant circularly polarized laser pulses through an optical medium. We assume $\sigma -$ pulses are transmitted for digital data input `0' and $\sigma +$ pulses for input `1'.  

\begin{center}
\begin{figure}
\includegraphics[width=\columnwidth]{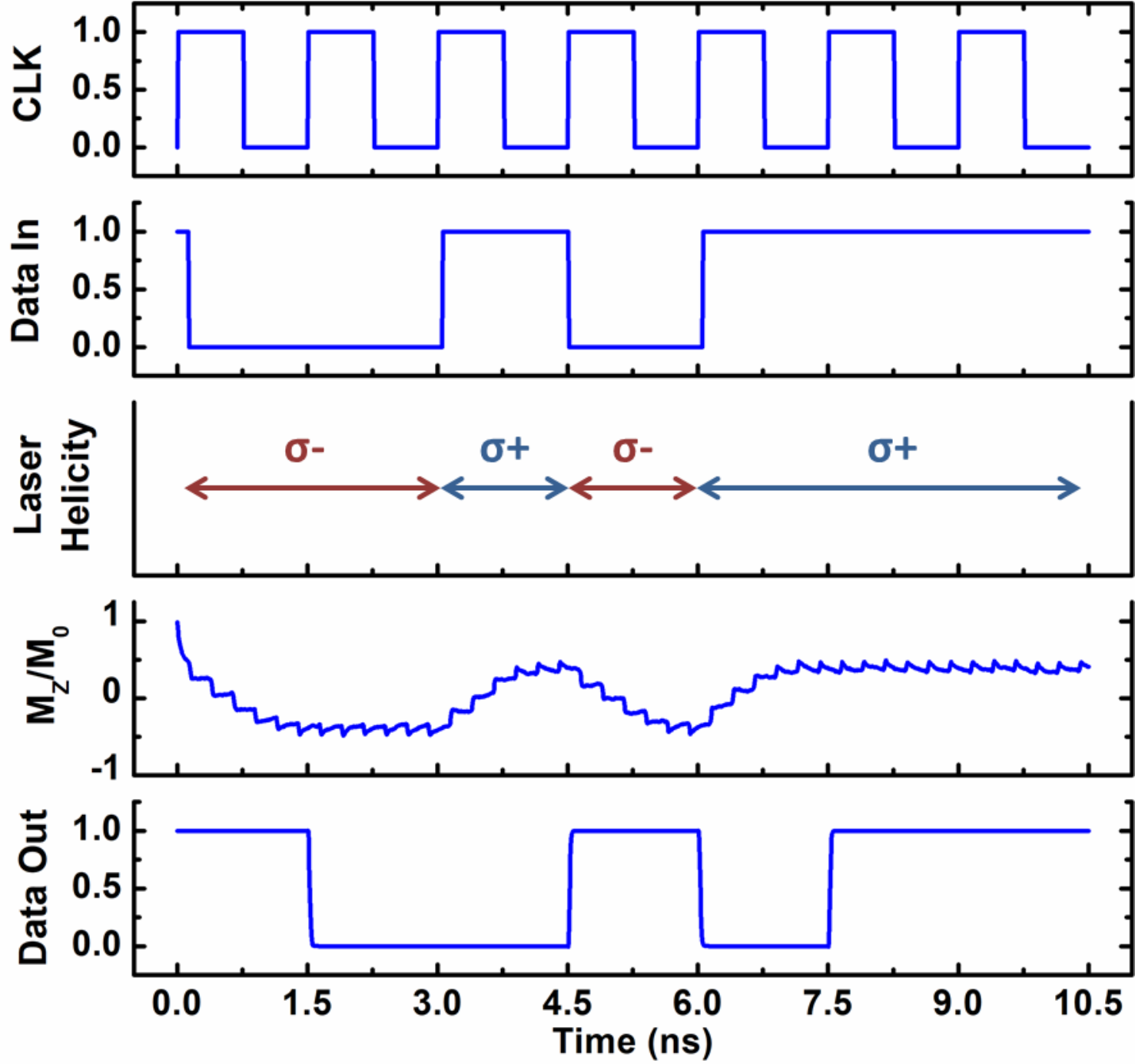}
\caption{Continuous operation of the interconnect circuit with a random input sequence}
\label{fig:circuit_operation}
\end{figure}
\end{center}

We show a sample operation in Fig.~\ref{fig:circuit_operation}. Here, continuous operation is shown for $7$ clock cycles with a random data input of `$0010111$'. We used $1.5$ $ns$ as the clock period to allow sufficient time for helicity induced magnetization reversal. We assume that the magnetization state of the Co/Pt free layer is initially pointing in the `up' direction ($M_{z}/M_{0} = +1$). In the first clock cycle, the input data is `0', which results in the transmission of $\sigma -$ pulses from the modulator. Since the free layer magnetization is initially in the `up' direction, the $\sigma -$ pulses reverse the magnetization towards `down' state. This is shown by the free layer magnetization ($M_{z}/M_{0}$) in Fig.~\ref{fig:circuit_operation}. At the end of the first cycle, the magnetization is read by activating the read voltage pulse and the output voltage goes to `0' following the clocked inverter (Fig.~\ref{fig:circuit_operation}). In the next cycle, the input data is again `0', which does not change the output magnetization. In the third cycle, the data input goes to `1' which results in $\sigma +$ laser pulse transmission. This results in the reversal of the magnetization towards `up' state as shown in Fig.~\ref{fig:circuit_operation}. At the end of this cycle, the data output goes to `1' in response to this magnetization reversal. The operation progresses in similar manner over the next cycles and data output follows the data input with one cycle latency (Fig.~\ref{fig:circuit_operation}). Next, we evaluate the performance of this optical receiver.  

\vspace{-6em}
\subsection{\label{sec:level5}Performance Evaluation}
The key feature of the proposed method is that the operation is simple which leads to an energy efficient performance. Using a SPICE simulation, we have evaluated the dissipated energy at the receiver to be $0.124$ $pJ/bit$. This is $\sim4\times$ lower than the required energy dissipation in the receiver using laser heat induced reversal~\cite{Azim2014}. The energy consumption is also $\sim5\times$ lower than the advanced Ge photodiode based receivers shown in~\cite{Zheng2010} and~\cite{Zheng2011}, which was reported to be the lowest among photodiode based receivers. The key limitation of our proposal, however, is the operating speed. This is because, magnetization reversal in ferromagnets through HDS is dictated by the cumulative action from multiple laser pulses. This is the major contrast in comparison with single-shot laser heat induced switching, where a single pulse can induce switching through ultrafast heating~\cite{Radu2011}. Hence, laser heat induced magnetization reversal in ferrimagnets is significantly faster than HDS in ferromagnets ($\sim5\times$ faster). However, optical receivers using laser heat induced magnetization reversal require the use of extra memory elements since there is no one-to-one correspondence between the laser pulse and magnetization state, which leads to the higher energy consumption. Moreover, as mentioned previously, laser heat induced magnetization reversal process applies primarily to ferrimagnets. Hence, the receiver in~\cite{Azim2014} requires the integration of ferrimagnet based MTJs, which creates additional design challenges. Our proposal only requires ferromagnetic MTJs, which is more desirable from a technology integration point of view. In spite of the slower operating speed, the proposed technique can be highly beneficial in situations where data needs to be transmitted over a very long distance at the lowest possible energy overhead with relaxed latency. 

\section{Conclusion}
To conclude, we have proposed helicity dependent switching of ferromagnets as an energy efficient process for optical-to-electrical signal conversion in optical interconnects. We developed a physics based model for HDS in ferromagnets and applied the model to develop a device to circuit level simulation framework. Our proposal shows the possibility of applying HDS to perform low power circuit operations.

\begin{acknowledgments}
This research was funded in part by C-SPIN, the center for spintronic materials, interfaces, and architecture, funded by DARPA and MARCO; the Semiconductor Research Corporation, the National Science Foundation, and the Vannevar Bush Faculty Fellows. T.~A.~Ostler gratefully acknowledges the support of the Marie Curie incoming BeIPD-COFUND fellowship program at the University of Li\`ege.
\end{acknowledgments}



\end{document}